\documentclass[showpacs,aps,pre,floatfix]{revtex4}
\usepackage{graphicx}

\begin{document}

\title{Non-universal dynamics of dimer growing interfaces}

\author{M. D. Grynberg} 

\affiliation{Departamento de F\'{\i}sica, Universidad Nacional de  
La Plata,(1900) La Plata, Argentina}

\begin{abstract}
A  finite temperature version of body-centered solid-on-solid growth 
models  involving attachment and detachment of dimers is discussed
in 1+1 dimensions. The dynamic exponent of the growing interface is 
studied numerically via the spectrum gap of the underlying evolution 
operator. The finite size scaling of the latter is found to be affected by
a standard surface tension term on which the growth rates depend. 
This non-universal aspect is also corroborated by the growth  behavior
observed  in large scale simulations.  By contrast,  the roughening 
exponent remains robust over wide temperature ranges.
\end{abstract}

\pacs{81.15.Aa, 05.10.Gg, 02.50.-r, 75.10.Jm}

\maketitle

\section{Introduction}

In studying statistical aspects of non-equilibrium surfaces the onset
of scaling regimes at both large time and length scales has enabled one
to characterize a vast body of growth processes in terms of universality 
classes [\onlinecite{Krug}].  In analogy to equilibrium phase transitions, 
there is consensus in that the late evolution
stages  of these processes are controlled by a set of scaling exponents 
stemming ultimately from  the symmetries and conservations laws of  the 
underlying growth rules. A basic quantity of interest investigated 
extensively  in this context concerns the roughness or surface width 
$W (L,t)$ developed by growth fluctuations at a given time $t$ when 
starting from an initially flat substrate of typical length $L$. Based on a 
wide  range of  theoretical and numerical studies it can be argued 
that $W$ scales as  [\onlinecite{Krug,Family}]
\begin{equation}
\label{DS}
W (L,t) = L^{\zeta}\, f ( t/L^z)\,,
\end{equation}
with a universal 
scaling function behaving as $f(x) \sim x^{\zeta/z}\,$ for $x \ll 1\,$, 
whereas for $x \gg 1\,$ it remains constant. Consequently, for $t \gg 
L^z\,$ the width saturates as $L^\zeta\,$ while growing as $t^{\zeta/z}$ 
in the thermodynamic limit. The roughening exponent $\zeta$ measures
the stationary dependence of the surface width on the typical substrate
size while $z$, frequently referred to as the dynamic exponent,
gives the fundamental scaling between length and time.

\begin{figure}[htbp]
\vskip -4.5cm
\centering
\includegraphics[width=0.9\textwidth]{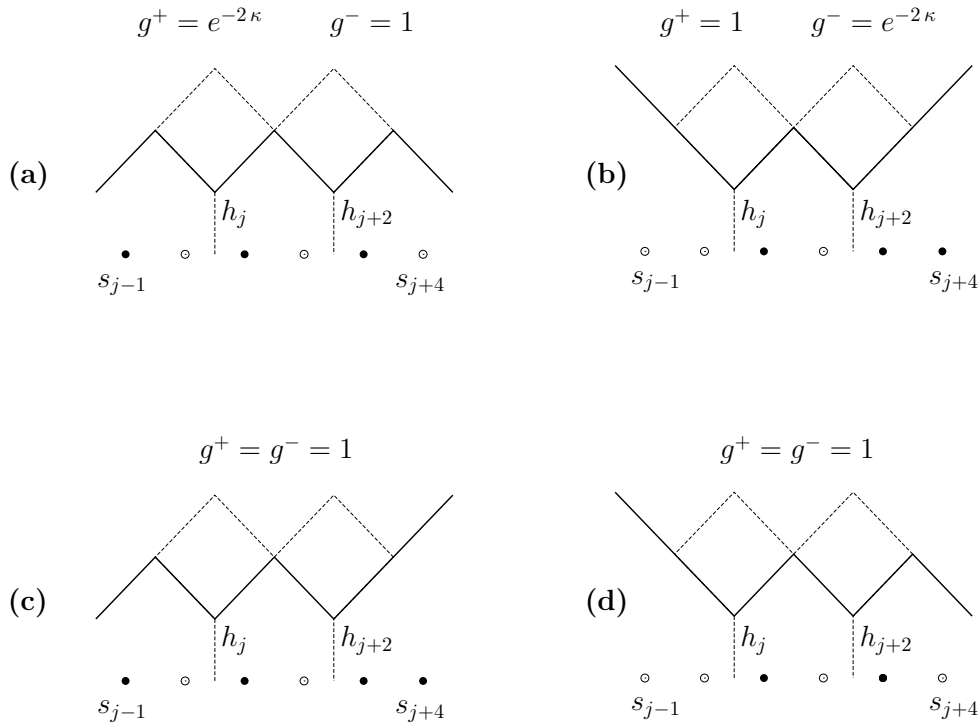}
\vskip -7.2cm
\caption{The four relevant deposition-evaporation cases and their 
respective rates $g^+, \, g^-$ for a one dimensional BCSOS dimer 
growing interface. Adding or removing  a dimer (denoted by dotted lines)
at columns $h_j, h_{j+2}$ can be viewed as flipping the spin-$\frac{1}{2}$ 
quartet $s_j, \,s_{j+1},\, s_{j+2}, \,s_{j+3}$.  The involved rates depend 
on the change of surface tension (\ref{tensionh}) which is in turn
determined by the spin (or slope) states of $s_{j-1}$ and $s_{j+4}$ 
[\,see Eqs.\,(\ref{ratesh}) and (\ref{rates})\,]. Each situation is depicted  
schematically from (a) to (d). }
\label{schematic}
\end{figure}

In this work we focus on rather unusual scaling properties studied in 
recent years both in one [\onlinecite{Nijs1,Hinrich,Gryn}] and two 
dimensional interfaces [\onlinecite{Nijs2}] by means of discrete models
of surface growth. For simplicity, here we consider a body-centered 
solid-on-solid (BCSOS) version of these [\onlinecite{Gryn}], limiting 
height differences between neighbors to $\pm 1$ rather than to
$0, \pm 1$, as in restricted SOS realizations [\onlinecite{Nijs1,Hinrich}].
Our basic kinetic steps (depicted schematically in Fig.\,\ref{schematic}), 
involve adsorption and desorption, possibly after recombination, of 
dimers only. Attempts of  desorption can take place whether or not the
selected pair of adjacent monomers arrived together. Therefore, the 
rule for evaporation allows for reconstitution of dimers, a crucial feature,
so their identity is not maintained during the growth process.  

It is important to note that  throughout the stochastic evolution the parity 
of  the number of monomers (eventually isolated) is conserved at each
height level of the surface. The implications of this `evenness' non local
constraint on the scaling exponents are far reaching 
[\onlinecite{Nijs1,Nijs2}], and in the one dimensional $(1d)$ case 
have been analyzed in terms of even visiting random walks.
In this latter representation, interface configurations are thought of as
Brownian paths whose locations (i.e. height levels) are visited an
even number of times  before the walk terminates on a given time 
interval (here playing the role of the  substrate length). In marked 
contrast to normal random walks, the evenness constraint 
introduces highly correlated movements giving rise to an 
anomalous (sub diffusive) mean square displacement, which in the 
interface language means a saturated width scaling {\it not} as $L^{1/2}$ 
but rather as $L^{1/3}$ [\onlinecite{Nijs1}]. Also, the dynamic exponents
$z\,$ obtained numerically for these [\onlinecite{Nijs1,Hinrich,Gryn}] and
related globally constrained systems [\onlinecite{Kim1,Kim2}]
are definitely different from those of usual monomer type interfaces, 
irrespective of the later evolving towards equilibrium or nonequilibrium 
stationary regimes (as exemplified respectively by the Edward-Wilkinson
[\onlinecite{EW}] (EW) and Kardar-Parisi-Zhang  [\onlinecite{KPZ}] (KPZ) 
universality classes). These anomalous aspects of non local constraints
should not be regarded as purely academic. In fact, in catalytic surface
processes the interplay between the substrate geometry  and the shape
of the intervening objects does matter. In particular, dimers  become
relevant in  the roughening dynamic of vicinal surfaces which only allows
deposition and evaporation of diatomic molecules [\onlinecite{Nijs2}].

To our knowledge, there are no available phenomenological 
equations of growth (e.g. KPZ, EW), suitable to describe interface 
fluctuations arising from deposition-evaporation of {\it composite} 
particles. The manner in which non localities could be incorporated in 
that continuum limit is by far not clear.  However, in probing  the 
robustness  of the dissociative dimer models referred to above some 
progress can be made by introducing a continuously tunable parameter
without affecting neither the symmetries nor the conservation laws of
their dynamics. Specifically, we consider a finite temperature extension
of $1d$ dimer growing interfaces, which for the ease of our numerical 
analysis (Sec. III\,A), is here taken under detailed balance conditions.
Following Ref.\,[\onlinecite{Amar}], this is readily  done through a 
standard surface tension term associated to configurational energy
scales which discourage the development of strong fluctuations, besides
those already prevented by both  BCSOS and evenness constraints. In 
spite of these  severe restrictions, they should not impede us to evaluate
scaling exponents, as equilibrium surfaces at finite temperatures are
always rough in $1d$. In the case of $1d$-monomer growing interfaces,
the surface tension is not relevant to drive them out of their universality 
classes, though interestingly in $2d$ it can change the nonlinear term 
sign of the KPZ equation thus inducing rough-to-rough transitions
[\onlinecite{Amar,Krug2}].  
Surprisingly, for dimers it will turn out that the combined effects of global 
constraints and surface tension entail significant changes in the $z$ 
dynamic exponent, suggesting rather a non-universal temperature
dependent value. In fact, non-universal aspects were already observed 
over wide temperature ranges in the monomer systems studied in 
Refs.  [\onlinecite{Amar,Krug2}], although they were ascribed to finite 
size effects which become particularly severe in nearing the equilibrium 
roughening transition temperature $T_R$ [\onlinecite{Chaikin}]. However 
as mentioned above, in our $1d$ case $T_R $ is strictly zero and in 
practice all equilibrium correlation lengths can be fairly bounded as long 
as temperatures are not taken too low. In this sense, it is worth 
mentioning that steep variations of $z$ will already appear within high 
temperature regimes (Sec. III). By contrast, the roughening exponent 
$\zeta$ remains robust as differences  with respect  to its vanishing 
tension limit $(\zeta \sim 1/3)$, will basically merge with our numerical 
errors. Since by definition $\zeta\,$ is a stationary index, it is just the 
nonequilibrium dynamics that is being strongly affected, as we shall see.

Turning to methodological issues, as is known the dynamic scaling 
hypothesis referred to in Eq.\,(\ref{DS}) is usually put forward to 
determine both $\zeta$ and $z$ in a jointly way by simulating the growth 
dynamics over different substrate sizes. With the aim of obtaining an 
independent (separate) evaluation of these exponents, in addition to this
standard procedure we will also exploit the known equivalence between 
BCSOS models and interacting gases of hard-core particles 
[\onlinecite{Meakin}]. Following the thread of ideas given in 
[\onlinecite{Robin}], we will recast the Metropolis operator that rules our 
growth simulations in terms of a quantum spin representation.
This latter lends itself more readily for a finite size scaling analysis
of the {\it gap} of the Metropolis operator which ultimately is related to 
the $z$ dynamic exponent. On the one hand this technique avoids the 
problem of dealing with long transient regimes though on the other
is limited severely by the affordable substrate sizes.  For now let us 
simply remark that already modest lengths are able to yield clear finite 
size trends over wide temperature ranges. To complement our 
approach, at low temperature regimes (where correlation lengths
exceed the sizes reachable by exact diagonalization), we will rely on 
numerical simulations of much larger systems which along with 
Eq.\,(\ref{DS}) will further support the non-universal picture.

The layout of this work is organized as follows. In Sec. II we construct the 
quantum spin analogy of the standard Metropolis dynamic and briefly 
touch upon symmetries and conservation laws.  By means of an ulterior 
non-unitary spin rotation, this results in a symmetric representation of the 
Metropolis operator.  This simplifies considerably the subsequent 
numerical analysis of Sec. III  in which the spectrum gap of this operator 
is obtained via standard recursive techniques [\onlinecite{Lanczos}].
The evaluation of dynamic exponents is then extended to low 
temperature regimes using standard Monte Carlo simulations. Finally, 
Sec. IV contains a  summarizing discussion along with some remarks on
extensions of this work.

\section{Dynamic and representations}

As usual,  the state of a solid-on-solid interface is represented by a set 
of single-valued functions $\{h_j (t)\}$ denoting height levels at positions 
$j = 1,\,...\,,L\,$  measured at a given time $t$ from a reference substrate 
of length $L$. As mentioned earlier, to prevent arbitrary bulk fluctuations 
we impose BCSOS constraints on these heights which hereafter are 
taken to satisfy  $h_{j+1} - h_j = \pm 1\,,\forall \,j,t\,$, along with periodic
boundary conditions (PBC). Growth or evaporation of the interface 
involves two particles (dimers) at the top of columns $h_j\,, h_{j+2}\,$
which, to comply with the above restrictions, ought both to be local
extrema of the evolving interface. More specifically, deposition 
(evaporation) events $h_j,\,h_{j+2} \to h_j + 2,\, h_{j+2} + 2\:\:  (\,h_j,\,
h_{j+2} \to h_j - 2,\, h_{j+2} - 2\,)$ can only occur at two consecutive 
local minima (maxima) of the heights set (see Fig.\,\ref{schematic}). 
We stress that evaporation takes place regardless if these maxima 
were created together or not, so dimers can dissociate. Also note that 
the number of heights at a given level preserves its parity throughout.

We want the  transition rates of these processes to depend on the 
surface tension $\sigma$ referred to in Sec. I.\, In turn, any model that 
associates energies to height  differences should provide a plausible 
description of $\sigma$.  Due to our BCSOS choice $(\Delta\,h = \pm 1)$, 
evidently the simplest form of $\sigma$ should assign an energy 
$\epsilon > 0$ to each double facet exposed  between columns $h_j$ 
and $h_{j+2}$. Since $h_{j+2} - h_j\ = 0,\pm 2\,$, this may be studied 
by defining
\begin{equation}
\label{tensionh}
\sigma = \frac{\epsilon}{2}\, \sum_j \left\vert\, h_{j+2} - h_j\,\right\vert\,,
\end{equation}
which simply counts the total number of double facets in a given 
interface configuration.   
Therefore, we can construct a  standard Metropolis process at 
temperature $T$ after introducing the following transition 
probability rates
\begin{equation}
\label{ratesh}
g^{\pm} ( \,h_j\, , \,h_{j+2}\, \to  \, h_j \pm 2\,, \,h_{j+2} \pm 2 ) = 
\min \left\{\,e^{-\Delta_j \sigma/T},\,1\right\}\,,
\end{equation}
where $\Delta_j \sigma$ is the change in surface tension upon
depositing $(g^+)$ or removing $(g^-)$ a dimer at $h_j,\,h_{j+2}$  
(henceforth, the Boltzmann constant $k_B$ is set equal to one). 
The four possible scenarios determining the values of these rates,
which by construction obey detailed balance (that is $g^{\pm} (\Delta
\sigma/T)/ g^{\pm} (-\Delta \sigma/T) = e^{-\Delta \sigma/T}\,)$,
are schematized in Fig.\,\ref{schematic}. Clearly, in the high 
temperature limit we recover the dimer model considered in 
[\onlinecite{Gryn}], whereas  $T \to 0^+$ serves to favor smooth
states over long transient regimes.

It is often more practical to work in terms of slopes rather than with 
interface heights, so in what follows we will employ the known mapping 
between BCSOS and hard core particle dynamics  [\onlinecite{Meakin}].
This correspondence is easily visualized in Fig.\,\ref{schematic} which
simply associates the height differences $h_j - h_{j-1} \equiv s_j$ to 
particles $(s_j = 1)$ or vacancies $(s_j = -1)$. Conversely, the interface
heights are obtained as $h_j = \sum_{n \le j} s_n\,$, modulo a constant
level. In particular, in this picture the surface tension reduces to the 
Ising  Hamiltonian
\begin{equation}
\label{tension}
\sigma = \frac{\epsilon}{2}\, \sum_j s_j\, s_{j+1}\,,
\end{equation}
up to an irrelevant constant; whereas after some straightforward 
manipulations, the square deviation of the instantaneous average height
 $\bar h$ of a particular slope configuration $\vert \,s \,\rangle$,  i.e. a 
given realization of  the  interface `width' $W^2 _{\vert \,s \,\rangle} 
\equiv \sum_j  \,( \,h_j - \bar h \,)^2  /L$, has the form
\begin{equation}
\label{width}
W^2_{\vert\, s \,\rangle }= \frac{L^2 - 1}{6\,L}\,+\,\frac{2}{L^2}\,
\sum_{i < j } \, i\, (L-j)\, s_i \, s_j\,.
\end{equation}
Creating (eliminating) a dimer now amounts to a backwards\, 
$\circ~\bullet~\circ~\bullet~\rightarrow~\bullet~\circ~\bullet~\circ$
(forwards\, $\bullet~\circ~\bullet~\circ~\rightarrow~\circ~\bullet~\circ~
\bullet$) move of two particles at a time, whereas from Eq.\,(\ref{tension})
it can be readily checked that in this representation the corresponding 
rates (\ref{ratesh}) just involve the neighboring states of the flipped 
quartet $s_j, s_{j+1}, s_{j+2},s_{j+3}$, namely
\begin{equation}
\label{rates}
g^{\pm}_j = \min \left\{\,e^{\pm \, \kappa\, \left (\,
s_{j+4} \,-\, s_{j-1}\,\right)}, 1 \,\right\}\,,
\end{equation}
with $\kappa \equiv \epsilon/T$ being from now on our inverse 
temperature parameter. Note also that under PBC the interface 
`magnetization' $\sum_j s_j$ vanishes at all evolution stages. 

In addition, some constants of motion can be immediately identified. 
Evidently, the dimer dynamic exchanges two particles between four
consecutive sites while changing the occupation of the involved next 
nearest neighbor locations by the same amount. If we think of these 
sites as being  part of a four-partite lattice $\Lambda = \Lambda_1 + ... 
+ \Lambda_4$  ($L/2$ even), hence upon defining  $S_{\alpha} \equiv
 \sum_{j \in \Lambda_{\alpha}} s_j$ as the magnetization of sublattice  
$\Lambda_{\alpha}$  it is clear that the set of dependent quantities 
$\{ \,(-1)^{\alpha} S_{\alpha} \,-\, (-1)^{\alpha'} S_{\alpha'}\,,\, \alpha,\, 
\alpha' = 1, ... , 4\,\}$ (of which only three are independent), is left 
invariant throughout. From a more fundamental point of view these 
conservations arise  ultimately from continuous  symmetries borne
by the Metropolis operator controlling the probabilities of our slope
states, and towards which we now turn.

\subsection{The Metropolis operator}

As is known, the evolution operator of a Markovian process of the kind 
discussed so far can be constructed generically as [\onlinecite{Kampen}] 
\begin{eqnarray}
\label{non-diagg}
\langle\,s'\,\vert\,M\,\vert\,s\,\rangle &=& -\,R(s \to s')
\hspace{0.4cm},\hspace{0.4cm} s \ne s'\,,\\
\label{conservation}
\langle\,s\,\vert\, M\,\vert\,s\,\rangle &=& \sum_{s'\ne s}\, R(s \to s')\,,
\end{eqnarray}
where $R ( s \to s') $  denotes the transition rate at which configuration 
$\vert s \rangle$ evolves to $\vert s' \rangle$ per unit time. At least 
formally, this enables one to derive all subsequent probability 
distributions $\vert P(t) \,\rangle \equiv \sum_s P (s,t)\, \vert s \rangle\,$ 
from the action of the evolution  operator on a given initial state, that is 
$\vert P(t) \,\rangle = e^{- M\,t} \vert P(0)\,\rangle$ [\onlinecite{Kampen}].
In our problem,  $R ( s \to s') = 1,\, e^{-2 \kappa}$ for all compatibles  
$\vert s \rangle,\,\vert s' \rangle$, and the specific form of $M$ can be 
readily found by interpreting the slope configurations $\vert s \rangle 
\equiv \vert s_1, ...\,, s_L \rangle$ as eigenstates of the $z$ component, 
say, of Pauli  matrices $\vec \sigma_1,\,...\,,\vec \sigma_L$  assigned to 
each  slope site. For instance, using spin-$\frac{1}{2}$ raising and 
lowering operators  $\sigma^+, \sigma^-\,$, the operational analog of 
Eq.\,(\ref{non-diagg}) will then read
\begin{equation}
\label{non-diag}
\sum_{s,s',\; s \ne s'}\!\!  M_{s',s}\; \vert s' \rangle
\,\langle s \vert = - \sum_j\,
\left(\, g^+_j\,A^{\dag}_j\,+\,g^-_j\,A_j\, \right)\,,
\end{equation}
where the adsorption (desorption) or double exchange operator 
$A^{\dag}_j $ $(A_j)$ acting on the $j$-quartet referred to above
is simply
\begin{equation}
\label{A}
A^{\dag}_j = \sigma^+_j\, \sigma^-_{j+1}\,
\sigma^+_{j+2}\,\sigma^-_{j+3}\,.
\end{equation}
Here, $g^{\pm}_j$ are thought of as diagonal operators in the
$\sigma^z$ representation and whose elements are identified with 
the rates of Eq.\,(\ref{rates}). In this regard, the ordering
of application in (\ref{non-diag}) is immaterial. As for the diagonal 
elements of Eq.\,(\ref{conservation}), needed for conservation of
probability, they basically count the number of ways 
in which a given configuration $\vert s \rangle$ can evolve to different
states $\vert s'\rangle$ by flipping an active quartet at a time. This can 
be properly  tracked down in terms of number operators $\hat n = 
\sigma^+ \sigma^- = (1+\sigma^z)/2$ and weighting each probed 
quartet with its corresponding rate ($g^{\pm}$). The counterpart
of  Eq.\,(\ref{conservation}) then becomes
\begin{eqnarray}
\label{diag}
\sum_s\, M_{s,s}\; \vert s \rangle \,
\langle s \vert &=& \sum_j \, g^+_j\, (1-\hat n_j)\, \hat n_{j+1}\,
(1-\hat n_{j+2}) \, \hat n_{j+3}\\
\nonumber
 &+& \sum_j \,g^-_j\, \hat n_j\, (1-\hat n_{j+1})\, \hat n_{j+2}\,
 (1-\hat n_{j+3})\,,
\end{eqnarray}
which along with Eq.\,(\ref{non-diag}) completes the form of our 
Metropolis operator. Taking into account the spin algebra  
$[\,\sigma^+_i\,,\,\sigma^-_j \,]  = \delta_{i,j}\,\sigma^z_j,\,$ 
$\{\,\sigma^+_j \,,\,\sigma^-_j \,\}_+\equiv 1,$ the former finally 
reduces to 
\begin{equation}
\label{non-hermitian}
M = \sum_j\,\left(\, g^+_j\, A^{\dag}_j\,+\, g^-_j \, A_j \,\right) 
\left(\, A^{\dag}_j\,+\, A_j\,-\,1\,\right)\,.
\end{equation}
By construction $M$ is a stochastic operator and therefore its 
ground state $\vert \Psi_0 \rangle$ has vanishing  eigenvalue and 
corresponds to the detailed balance solution of the problem,  
i.e. $\vert \Psi_0 \rangle \propto \sum_s e^{-\frac{\kappa}{2} 
\sigma _{\{s\}} } \vert s \rangle$. Instead, its {\it left} ground
state $\langle \tilde \psi \vert$ is an equally weighted linear
combination of all reachable $\langle s \vert$ (note that $M$ is a non 
hermitian operator whose columns add up to zero). With the aid of this 
left state and starting from an initial probability distribution  $\vert P (0) 
\rangle$, typical quantities of interest, such as the interface width $W^2$, 
are calculated as $\langle \tilde \psi \vert \,\hat{\cal W}\, e^{-M t} \vert 
P (0) \rangle$  [\onlinecite{Kampen}]. Here, the `width operator' $\hat 
{\cal W}$ is obtained by promoting the slopes of  Eq.\,(\ref{width}) to 
$\sigma^z$ matrices.

Despite the apparent simplicity of our Metropolis operator, the $A$'s
above can not be associated to elementary excitations of any kind 
and exact analytic treatments may seem unlikely.  Nevertheless,  
Eq.\,(\ref{non-hermitian}) will permit some numerical progress 
on finite size systems  after considering a simple transformation
to be discussed later on in Sec. II\,C. Before that and for the
sake of completeness, we pause to digress 
briefly about symmetries and conservation laws of $M$.

\subsection{Excursus: constants of motion}

Here we follow  Refs.\,[\onlinecite{BGS,DB}]  in closely related 
processes. Recalling  that under a rotation by an angle $\theta$ around 
the  $z$-direction $\sigma^{\pm}$ transform as $e^{ \pm \rm i \theta} 
\sigma^{\pm}$, we can therefore choose angles $\theta_{\alpha}$ for all 
spins in each sublattice $\Lambda_{\alpha}$ such that  Eq.\,(\ref{A})
[\,and obviously (\ref{diag}) ] is left invariant. Clearly, this is  the case of 
\begin{equation}
\label{angles}
\sum_{\alpha} (-1)^{\alpha} \,\theta_{\alpha} = 0\,.
\end{equation}
On the other hand, the infinitesimal generator of this transformation is
$S = \sum_{\alpha} (-1)^{\alpha}\, \theta_{\alpha} \,S^z_{\alpha}\,$, 
with $S^z_{\alpha} = \sum_{j \in \Lambda_{\alpha}} \sigma^z_j$. Since 
$M = e^{\,{\rm i} S /2 } M  e^{-{\rm i} S /2 }$, then $[ M\,,\,S] = 0\,$, 
and therefore $S$ is preserved by $M$. But from the constraint
(\ref{angles}) it follows that $S$ can be rewritten in terms
of three independent angles, that is $S = \sum_{\alpha \ne \alpha'}
\theta_{\alpha} \,[\,(-1)^{\alpha} S^z_{\alpha} \,-\, 
(-1)^{\alpha'} S^z_{\alpha'}]$ from which one recovers the three 
conserved quantities identified before on more intuitive grounds.

These continuous symmetries entail a number of invariant  subspaces
growing at most as $L^3$, which however by no means exhaust all 
possibilities. For instance, it is straightforward  to see that already the 
number of jammed configurations (i.e. states that can not evolve further),
grows {\it exponentially} with the system size [\onlinecite{BGS}]. This 
unusual proliferation of invariant states should be the consequence
of a much higher symmetry of $M$. Although its explicit operational form
might be difficult to figure out, we can nevertheless follow 
Ref.\,[\onlinecite{DB}]  and construct an exponential number of  
dynamically disjoint sectors, either jammed or unjammed, regardless of
the value of $\kappa$. To this end, one defines a reduction rule by 
looking at the occurrence of groups of active quartets in a given
configuration $\vert s \rangle$. Each occurrence, if any, is deleted so
the length of the remaining object is reduced in 4-bits per deletion. This
procedure is applied recursively until one is left with a string that can not
be further reduced, i.e. an {\it irreducible string} $I \{s\} $. In turn, the
result is unique irrespective of  the order of deletion. To mention only a 
few examples: $I \{\vert\uparrow\downarrow\uparrow\downarrow\uparrow
\downarrow\uparrow\downarrow\downarrow\uparrow\,\rangle\} = 
\;\downarrow\uparrow\,$ (either in one or two steps);  the flat interface or
antiferro state yields a null string, whereas any jammed configuration is
already an irreducible string of length $L$. The key issue to bear in mind
is that two states  $\vert s \rangle$ and $\vert s' \rangle$ belong to the
same $M$-subspace $\Leftrightarrow I \{ s \} = I \{ s' \}$ [\onlinecite{DB}]. 
So, this non-local construct picks out both  the length  ${\cal L} = L - 4k$
and the sequence of the irreducible string's elements (their combinations 
growing exponentially in ${\cal L}$), and ultimately defines the constant
of motion under which the $M$-dynamics take place.

That being said, from now on we shall content ourselves with studying
just the null string subspace selected by initially flat conditions, for the 
most part quite natural in the context of growing interfaces. Note also 
that the equilibrium properties of the $1d$-Hamiltonian (\ref{tension}) are 
neither analytically simple to evaluate (e.g. Eq.\,(\ref{width}), not even for 
$\kappa = 0$), as the ensemble of  averaged states must be consistent 
not only with $S^z = 0\,$ (PBC) but with a vanishing irreducible string as
well (totally unjammed conditions), which rules out an exponential 
number of states.

\subsection{Symmetric representation}

Returning to the discussion of Sec.\,IIA, we may make some progress on
the numerical analysis of $M$ by performing a similarity  transformation
so as to map this operator into an hermitian matrix.  This is feasible 
because detailed balance in rates (\ref{rates}) ensures the existence 
of a representation in which the evolution operator is self adjoint
[\onlinecite{Kampen}]. For this purpose, it suffices to consider a diagonal 
transformation alike the one discussed in Sec. II\,B but using pure 
imaginary angles instead. Specifically, we rotate each $j$th spin around 
the $z$-direction by a site dependent angle (field operator)
\begin{equation}
\label{angles2}
\varphi_j =  \frac{{\rm i}\,\kappa}{2}\, \left(\, \sigma^z_{j-1} +
 \sigma^z_{j+1} \,\right)\,,
\end{equation}
by means of the nonunitary similarity transformation $U = e^{ - {\rm i} 
\sum_j \varphi_j \sigma^z_j/2}$. Under this rotation $\sigma^{\pm}_j \to  
e^{\pm \frac{\kappa}{2}  (\sigma^z_{j-1} +\, \sigma^z_{j+1}) }
\sigma^{\pm}_j$, so it is simple to check that the double hopping 
operators of  Eq.\,(\ref{A}) transform as
\begin{eqnarray}
\nonumber
U\,A^{\dag}_j\,U^{-1} &=& e^{ \frac{\kappa}{2} \,\left( \sigma^z_{j-1} 
\,- \;\sigma^z_{j+4} \right) } A^{\dag}_j\,,\\
U\,A_j\,U^{-1} &=&  e^{- \frac{\kappa}{2} \,\left( \sigma^z_{j-1} 
\,- \; \sigma^z_{j+4}  \right) } A_j\,.
\end{eqnarray}
This introduces new diagonal operators $\hat d_j$ in the  $\sigma^z$  
representation, that like the $g^{\pm}_j$ operators commute
with $A^{\dag}_j, A_j\,$, and in terms of which the off diagonal part  of 
$M$ becomes symmetric. More specifically, this symmetrization is 
produced by defining
\begin{equation}
\hat d_j = e^{ \frac{\kappa}{2} \,\left( \sigma^z_{j-1} \,- \; 
\sigma^z_{j+4} \right) }\, g^+_j 
=  e^{ -\frac{\kappa}{2} \,\left( \sigma^z_{j-1} \,- \; 
\sigma^z_{j+4} \right) }\, g^-_j \,,
\end{equation}
their diagonal elements being $e^{- \frac{\kappa}{2}\, \left\vert s_{j-1}\,-
\,s_{j+4}\right\vert}\,$. As a result,  Eq.\,(\ref{non-diag}) is transformed 
into  $-\sum_j \hat d_j ( \,A^{\dag}_j  + A_j )\,$ while Eq.\,(\ref{diag}) is left 
unchanged, so the rotated Metropolis  operator $H = U M U^{-1}$ can 
be finally cast in the symmetric form 
\begin{equation}
\label{symmetric}
H = \sum_j\,\left(\,  A^{\dag}_j\,+ \,A_j \,\right) 
\left(\, g^-_j\, A^{\dag}_j\,+\, g^+_j\, A_j\,-\,\hat d_j\,\right)\,.
\end{equation}

Consequently, the time dependent probability distribution turns out to
be a superposition of orthogonal eigenlevels $\vert \psi_{\lambda \ne 0}
\rangle$ with {\it real} eigenvalues $\lambda > 0$ of $H$, each having
typical lifetimes $1/\lambda$.
In particular, the ground state $\vert \psi_0 \rangle$  has eigenvalue 
$\lambda = 0$, and is just the transformed Boltzmann distribution 
$\vert \Psi_0 \rangle$ referred to above, i.e. $\vert \psi_0 \rangle = 
U \vert \Psi_0 \rangle \propto \sum_s  e^{-\frac{\kappa}{4} \sigma _{\{s\}}}
\vert s \rangle$.  Since left and right levels now coincide, it is thereby a 
simple matter to check that in the symmetric representation
the dynamic of any diagonal observable, say the interface width $\hat 
{\cal W}_L = \frac{2}{L^2}\, \sum_{i < j } \, i\, (L-j)\, \sigma^z_i \, 
\sigma^z_j\,$ (in turn invariant under $U$), can be written as
\begin{equation}
W^2 (L,t) =  W^2_{eq} \,+\,  \sum_{\lambda_{_L} > 0}\, 
e^{-\lambda_{_L} \! t}\, \langle \, \psi_0\,\vert \,\hat {\cal W}_L \,\vert \,
\psi_{\lambda_{_L}}\,\rangle\,\langle \, \psi_0\,\vert P'(0)\,\rangle\,,
\end{equation}
where  $W^2_{eq}$ is the saturation width reached at equilibrium,
whereas $\vert P'(0) \rangle = U\, \vert P(0)\rangle$ denotes the
transformed initial distribution. From here we see that if the spectrum gap
vanishes as $1/L^z$, then a  finite size scaling analysis of the first excited 
levels $\lambda_L$ will provide the dynamic $z$-exponent ruling over
the late roughening stages referred to in Sec. I.

It is worth pointing out that the discussion presented so far can be
readily extended to include monomers $(m=1)$,  trimers $(m=3)$, etc.\,, 
so long  as the operators involved in Eq.\,(\ref{symmetric}) are 
reinterpreted as
\begin{eqnarray}
\nonumber
A^{\dag}_j &=& \prod_{i=1}^m \sigma^+_{j + 2 i - 2}\: \sigma^-_{j+2i -1}\,,
\\
\label{kmers}
g_j^{\pm} &=& \min \left\{\,e^{\pm \, \kappa\, \left (\,s_{j+2 m} \,-\, s_{j-1}\,
\right)}, 1 \,\right\}\,,
\\
\nonumber
\hat d_j &=& e^{- \frac{\kappa}{2}\, \left\vert s_{j-1}\,-\,s_{j+2 m}
\right\vert}\,.
\end{eqnarray}
In particular, for monomers with no surface tension ($\kappa = 0$)
the evolution operator reduces to the fully isotropic Heisenberg 
ferromagnet, thus recovering the usual EW dynamic exponent $z=2$.
Among other numerical aspects, in  what follows we shall focus on the 
evaluation of this quantity at $\kappa \ge 0$ for both $m = 1$ and $2$.

\section{Numerical results}

The above ideas provide an alternative manner to evaluate dynamic 
exponents, independently of those obtained by the application of the 
dynamic scaling hypothesis (\ref{DS}). Thus, we first explore the 
consequences arising from the exact diagonalization of 
Eq.\,(\ref{symmetric}) in small systems, and then go on to corroborate
them over larger length scales [via Eq.(\ref{DS})\,] using standard 
Monte Carlo simulations. In addition, these latter will complement the 
evaluation of $z$ in low temperature regimes where, as we shall see,
resorting to small lattice sizes might become inadequate.

\subsection{Scaling the gap}

To analyze our stochastic matrix we first obtained its null string basis 
using dimers on  rings of sizes $L = 4 k$. This was easily implemented by
applying $H$ (or alternatively, $M$) to either of  the two antiferro states
and keeping proper track of the new generated configurations.  By
iterating this procedure with those new states for which $H$ was not 
previously applied, the whole null-string subspace was finally  expanded.
The total number of states so found is of course independent of 
$\kappa$, and as expected (see Sec. II\,B), grows slower than the 
monomer space dimensionality  ${L \choose L/2} \propto 2^L$. More 
specifically, this dimension seems to increase as  $\approx 1.6(7)^L$, 
at least for the sizes at hand, which in turn allowed us to explore rings
of up to 32 sites [\onlinecite{SOS}].

\begin{figure}[htbp]
\vskip -2cm
\centering
\includegraphics[width=0.5\textwidth]{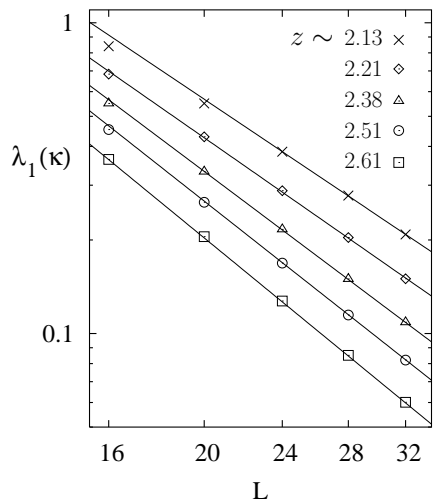}
\vskip -3cm
\caption{Finite size behavior of the first excited level $\lambda_1$
of the Metropolis operator (\ref{symmetric}). The inverse temperature
$\kappa$ decreases from top to bottom, each symbol standing 
respectively for $\kappa = 1,\ \frac{2}{3},\, \frac{2}{5},\,\frac{1}{5},\,0$. 
The dynamic exponents $z$ are read off from the slopes of  the
fitting lines. }
\label{gaps}
\end{figure}

Once having identified the null string configurations, we proceeded to 
evaluate exactly the low lying levels of $H$  (in principle, just the first 
excited  will do), via a recursion type Lanczos algorithm 
[\onlinecite{Lanczos}]. Starting that recursion from a random linear 
combination of null  strings but chosen orthogonal to the Boltzmann-type 
distribution $\vert \psi_0 \rangle\,$ referred to above, we then obtained 
the finite size behavior of the dimer gap, i.e. of $\lambda_1$. This is 
shown in Fig.\,\ref{gaps}  for several  temperatures within the range 
$0 \le \kappa \le 1\,$ indicating a gap decrease $\propto L^{-z}\,$,
however  notice that the data do not fall into parallel straight lines.
In an attempt to standardize this situation we used scalings of the form 
$\lambda_1 (L) = A_{\kappa} L^{-z_1} + B_{\kappa}  L ^{-z_2}$, but a 
large amount of uncertainty in both  $z_1$ and  $z_2$ raised doubts 
about the adequacy of such a procedure. Also, logarithmic corrections 
were attempted but no evidence supporting these latter were found. 
Thus, in principle we are led to suggest a plain power law decay 
although  with a {\it non-universal} temperature dependent dynamic 
exponent $z = z\,(\kappa)$. In particular, $z\,(0) \sim 2.6(1)$ is in fair 
agreement with the value obtained in Ref.\,[\onlinecite{Gryn}] by 
standard simulations of dimer interfaces without surface tension. 

\begin{figure}[htbp]
\vskip -2.5cm
\centering
\includegraphics[width=0.5\textwidth]{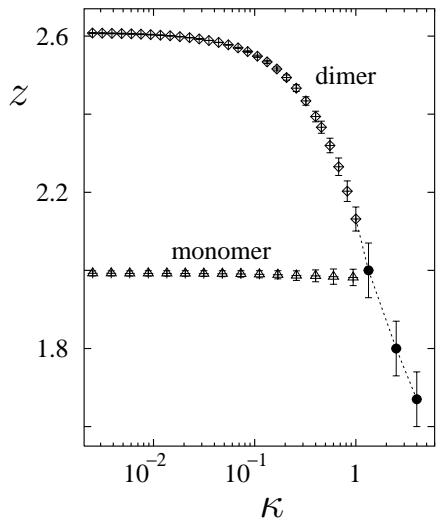}
\vskip -3cm
\caption{Non-universal exponents of the dimer dynamics. Rhomboids
indicate the results obtained from finite size scaling extrapolations of 
Metropolis gaps (see Fig.\,\ref{gaps} ).  For comparison, the triangles 
exhibit  these exponents in the monomer case (close to $z \approx 2$). 
Filled circles (joined by dotted lines) denote the $z$ values arising from 
the dynamic scaling hypothesis (\ref{DS}) applied to much larger systems 
at lower temperatures.}
\label{nonuniversal}
\end{figure}

A slightly improved estimation of $z$ can be made by defining an 
effective dynamic exponent 
\begin{equation}
z_L = \frac{\ln\left[ \,\lambda_1 (L-4) / \lambda_1 (L) \,\right]}
{\ln \left[ \,L / (L-4) \,\right] }\,,
\end{equation}
and then extrapolating $z_L$ to $L \to \infty$ for a given $\kappa$.
The results of this are exhibited in Fig.\,\ref{nonuniversal}, which for 
comparison also displays the corresponding monomer dynamic 
exponents. These latter were derived using Eq.\,(\ref{kmers}) 
for $m = 1$ along with a similar numerical analysis but employing 
$L = 2 k \, (\le 24)$ instead. Clearly, an EW behavior characterized by 
the Heisenberg exponent $z = 2$ mentioned a little earlier can be 
discerned in monomer interfaces, as opposed to dimer exponents 
which evidently are non-universal, at least if we are to judge by their 
$\sim 20 \%$ variation between  $\kappa \approx 0$ and 1. Although it
 is true that size effects increase monotonically our error margins with 
$\kappa$, they are nevertheless fairly bounded within the range 
inspected (see also Fig.\,\ref{gaps}).

\begin{figure}[htbp]
\vskip 0.5cm
\centering
\includegraphics[width=0.35\textwidth]{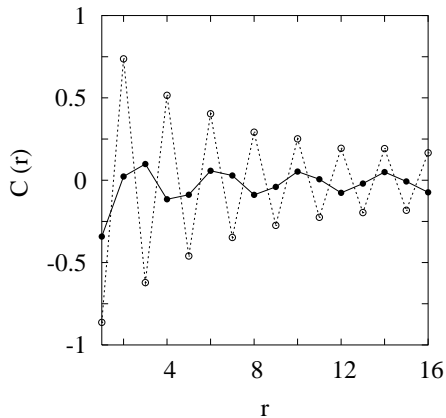}
\caption{Pair correlations in the ground state of operator 
(\ref{symmetric}) for 32 spins. Filled circles ($\kappa = 0$) are 
representative of high temperature regimes (which are quantitatively 
similar up to $\kappa \approx 1\,$). Above $\kappa = 2$ (open circles),
the underlying  correlation lengths  become rapidly comparable to the
lattice size.}
\label{correlations}
\end{figure}

A measure of these size effects is provided by the equilibrium 
correlation length of the associated Ising antiferromagnet  appearing 
in Eq.\,(\ref{tension}). If this length becomes comparable to our available
sizes, particularly at low temperatures, then the asymptotic dynamics 
will  be distorted on approaching equilibrium as  the average 
antiferromagnetic domain sizes, representing active regions in the 
interface, will be cut off by $L$. This situation is illustrated in 
Fig.\,\ref{correlations} where we show the pair correlations $C (r) = 
\frac{1}{L}\sum_j \langle \sigma^z_j \,\sigma^z_{j+r} \rangle$ evaluated 
in the ground state or equilibrium distribution of $H$ [as stressed above,
notice that analytic treatments are difficult even in this simpler case 
because of the null string constraint imposed on Eq.\,(\ref{tension})\,]. In 
between  $0 \le \kappa \alt 1\,$, traces of antiferromagnetic short range 
order are nearly smeared out and correlation lengths become small. 
However, above $\kappa  \approx 2$ they rapidly grow up and 
eventually get comparable to our maximum sizes, so precluding further 
analyses of the gap (which in fact comes out to be almost size 
independent). Thus, to complement the results obtained so far and 
check whether non-universal exponents actually extend down to low 
temperatures regimes, we finally turn to the dynamic scaling hypothesis
and simulations using larger substrates. 

\subsection{Simulations}

\begin{figure}[htbp]
\vskip -2cm
\centering
\includegraphics[width=0.5\textwidth]{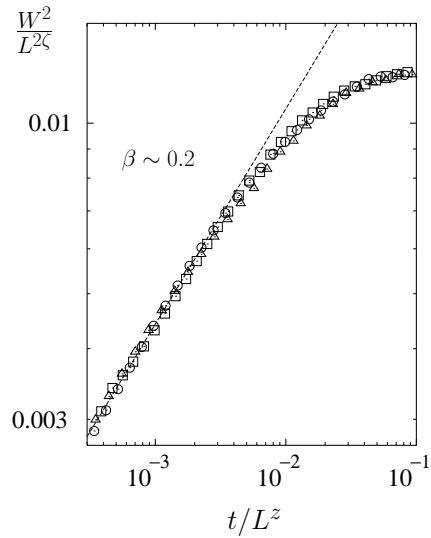}
\vskip -3cm
\caption{Dynamic scaling of the dimer interface width [\,Eq.\,(\ref{DS})\,]
taking $\kappa = 4$. Sizes $L = 2^{12}, \,2^{11}$ and $2^{10}$ are 
denoted respectively by triangles, circles and squares. The data collapse
was attained upon setting $\zeta \approx 0.3(4)$ and $z \approx 1.6(7)$. 
The dashed line is fitted with slope $2 \beta = 2\, \zeta/z$. }
\label{dynamic}
\end{figure}

Following the Metropolis rules referred to in Sec. II, we evolved initially
flat interfaces with $L = 2^{10}, 2^{11}$ and $2^{12}$ heights until 
reaching their stationary states. After a sequence of $L$ update attempts
at random locations, the timescale was increased in one unit, i.e. 
$t \to t+1$, irrespective of these attempts being successful or not. 
Measurements of $W (t)$ were carried out for $\kappa = 1.3,\, 2.5,\,4\,$ 
and were averaged  typically over $10^4$ independent histories. In 
Fig.\,\ref{dynamic} we display one of the characteristic scaling curves 
obtained using Eq.\,(\ref{DS}) for $\kappa = 4\,$ There, the data collapse 
was attained by setting roughening exponents $\zeta \approx 0.3(4)$ 
which are practically common to all temperatures studied (see also 
Fig.\,\ref{saturation} below). By contrast, this is not the case of the  $z$ 
exponents which, in line with the results of Sec. III\,A, are severely 
altered by $\kappa$. Although their precise values are blurred by our
not too sensitive collapse conditions, nevertheless they do follow the
non-universal trend already found with our gap analysis, as can be seen 
in Fig.\,\ref{nonuniversal}.

\begin{figure}[htbp]
\vskip -2.5cm
\centering
\includegraphics[width=0.53\textwidth]{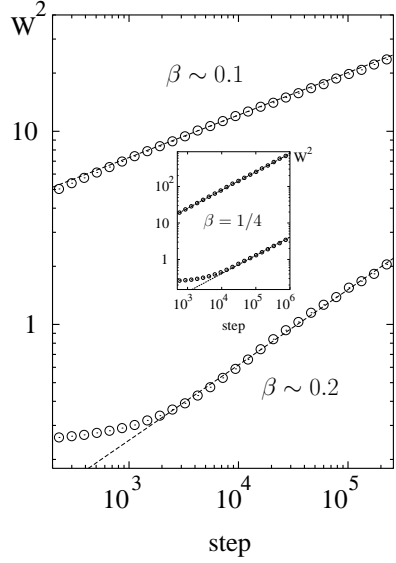}
\vskip -3.5cm
\caption{Growth of interface width for $10^6$ sites
at $\kappa = 0$ (upper curve), and $\kappa = 6$. Dashed lines
are fitted with slopes $2\,\beta$. The inset displays the corresponding
situations for monomer growing interfaces.}
\label{evolution}
\end{figure}

To corroborate further the validity of this claim, we also conducted 
simulations in much bigger scales measuring  directly the growth 
exponent $\beta \equiv \zeta/z$. The reader's attention is now directed 
to Fig.\,\ref{evolution} where the width evolution is contrasted at high and 
low  temperature regimes in substrates of $10^6$ sites.  As expected, 
non-universal aspects show up: after averaging over $\sim 40$ 
histories, clearly two rather different $\beta$ exponents emerge and 
hold for at least two decades.  On the other hand, using the universal 
roughening exponent $\zeta \approx 1/3$ already identified (see further
estimations below), we thus obtain values  of $z$ consistent with those 
previously encountered in smaller systems. For comparison, the inset
of  Fig.\,\ref{evolution} also shows the typical EW $\beta$-values of the 
corresponding monomer cases  which, alike their dynamical exponents
in Fig.\,\ref{nonuniversal}, remain robust under wide temperature 
intervals.

In respect of the roughening exponents, we finally considered the 
saturation or equilibrium widths $W_{eq}$ of a variety of substrate sizes 
subject to $\kappa = 0,\,1,\,2\,$ and 4.  The employed relaxation times 
$\propto L^{z (\kappa)}$, range from $2 \times 10^6$ to $10^5$ Monte 
Carlo steps for the largest cases and, as expected, decrease 
monotonically with $\kappa$. Our results are displayed in 
Fig.\,\ref{saturation}, clearly suggesting a common value of $\zeta$. 
Due to the pair correlations involved in $W_{eq}$ [\,see 
Eq.\,(\ref{width})\,], here size effects are also more noticeable at low 
temperatures. Nonetheless, a simple numerical fit of both amplitudes 
and slopes indicates that most of our data ($L \agt 200$) can be 
accounted for by the parametrization
\begin{equation}
W_{eq} \approx 0.4(2) \; e^{-\kappa\,\zeta} \, L^{\zeta}\,,
\label{param}
\end{equation}
with $\zeta \approx 0.3(2)$. This means that in equilibrium the interface
becomes actually rough so long as $\, \ln L \gg \kappa$ is held in the 
thermodynamic limit. 

\begin{figure}[htbp]
\vskip -2.2cm
\centering
\includegraphics[width=0.51\textwidth]{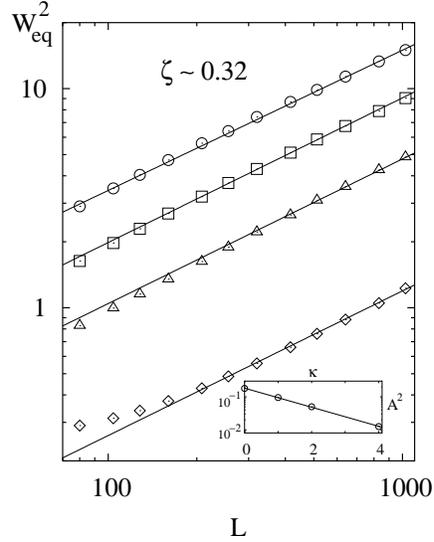}
\vskip -3cm
\caption{Finite size behavior of the saturation width. The symbols stand
in turn for $\kappa = 0$ (circles), $\kappa = 1$ (squares), $\kappa = 2$ 
(triangles) and, $\kappa = 4$ (rhomboids). Solid lines display a common 
slope $2\,\zeta$. As is shown in the inset, the amplitudes of these latter 
exhibit an exponential decay with slope $-2 \,\zeta$ [\,see 
Eq.\,(\ref{param})\,].}
\label{saturation}
\end{figure}

\section{Concluding discussion}

To summarize, we have studied numerically the dynamics of dimer
growing interfaces at finite temperatures using two independent 
procedures. The first one analyzes the spectrum gap of the evolution 
operator [\,Eq.\,(\ref{symmetric})\,] by exact diagonalization of small 
systems, thus picking out dynamic exponents in a direct manner. Clear 
finite size trends were obtained in the range $0 \le \kappa \le 1\,$ 
(Fig.\,\ref{gaps}), and fairly bounded extrapolations were derived for
$z$ (Fig.\,\ref{nonuniversal}). Although the non-local symmetries (i.e. 
irreducible strings of Sec. II\,B) of our stochastic operators are unaffected
by surface tensions, surprisingly the $z$ exponents are non-universal, 
being dependent on $\kappa$. Yet, a theoretical interpretation of such 
puzzling behavior remains quite open. This is in marked contrast with the
dynamics of monomers interfaces, as their tensions do not take over
neither the EW nor the KPZ universality classes, at least in $1+1$ 
dimensions [\onlinecite{Amar}].

Secondly, using the standard scaling hypothesis [\,Eq.\,(\ref{DS}), 
Fig.\,\ref{dynamic}] we checked out these findings under lower
temperature regimes where correlation lengths become larger than
our maximum diagonalizable sizes (Fig.\,\ref{correlations}). Despite the
limited precision of this method for $\kappa > 0$, our results confirmed
the non-universal tendency observed in Sec. III\,A . In turn, 
measurements of growth exponents in much larger substrates 
(Fig.\,\ref{evolution}) further validated the monotonic decrease of 
$z (\kappa)$. 

As for the roughening exponents (Fig.\,\ref{saturation}), in all studied
cases with $\kappa > 0$ the global constraint referred to in  Sec. I and
further examined in terms of irreducible strings  [\onlinecite{BGS,DB}],
led to anomalous motions of a rather unconventional type ($\zeta
\approx 1/3$), as compared to Levi flights and other restricted random
paths [\onlinecite{BN}]. More specifically, they are consistent with those
of even visiting random walks [\onlinecite{Nijs1}] and not comprehensible
in terms of  EW or diffusive interfaces (e.g. monomers), which are 
definitely rougher. Under surface tension the range of correlations so 
introduced in the associated walk is finite, and therefore the scaling of its
width must remain unchanged (consult Ref.\,[\onlinecite{BN}]\,), though in
line with Eq.\,(\ref{param}), its proportionality constant might depend on
the precise form of these correlations. To endow further this robustness
of $\zeta$, it would be interesting to elucidate whether the analogy of 
non-interacting electrons moving in a random medium studied in 
Ref.\,[\onlinecite{Nijs1}] could be extended to the finite temperature
interfaces (walks) investigated here. 

Other pending issues of interest concern starting the growth process 
from more general initial conditions (i.e. not in the null string sector),
capable of modifying asymptotic regimes [\onlinecite{Gryn}], as well as
considering biased dynamics (without detailed balance) such as those
analyzed in Refs.\,[\onlinecite{Amar,Krug2}]. In principle, the first situation
could also be studied with the methodology of Sec. II\,A; however for the
second one the similarity transformation of Sec. II\,C is no longer useful
and the unsymmetrization of the Lanczos recursion would be inevitable
[\onlinecite{Lanczos}].  Finally, $2+1$ dimensional generalizations of this 
study could shed light on the combined role that dimer dynamics and 
substrate geometry might have in catalytic processes. Whether or not 
non-universal aspects would also emerge there under surface tension,
deserves further investigations.

\section*{Acknowledgments}

The author is grateful to R. B. Stinchcombe for helpful observations and
correspondence. Support of CONICET, Argentina, under grants PIP 
5037 and PICT ANCYPT 20350, is acknowledged.



\begin{thebibliography}{99}
\bibitem{SOS} Had the restricted SOS constraint $\Delta h = 0, \pm 1\,$
been chosen, it would immaterially complicate the numerical analysis  
demanding instead spin-$1$ operators.
\end{thebibliography}
\end{document}